\documentclass[a4paper,11pt]{article}
\pdfoutput=1 

\usepackage{jcappub} 

\usepackage[T1]{fontenc} 

\usepackage{hyperref}
\usepackage{multirow}
\usepackage{graphicx}
\usepackage{dcolumn}
\usepackage{bm}
\graphicspath{{./figures/}}

\title{Interpretation of multi-TeV photons from GRB221009A}

\author[1]{Ali Baktash,}
\author[1]{Dieter Horns\note{Corresponding author.}}
\author[2]{and Manuel Meyer\note{Now at CP3-Origins, University of Southern Denmark, Campusvej 55, DK-5230 Odense M, Denmark}}
\affiliation{Universität Hamburg, Luruper Chaussee 149, D-22761 Hamburg, Germany}
\emailAdd{ali.baktash@uni-hamburg.de}
\emailAdd{dieter.horns@uni-hamburg.de}
\emailAdd{manuel.meyer@uni-hamburg.de}
\arxivnumber{2210.07172}
\abstract{
The nearby GRB221009A at redshift $z=0.1505$ has been observed up to a maximum 
energy of 18 TeV with the LHAASO air shower array. 
The expected optical 
depth for a photon with energy $E_\gamma=18$~TeV 
varies between 9.4 and 27.1 according to existing models of
the extra-galactic background light (EBL) in the relevant mid infra-red range. The  resulting suppression of the flux in several EBL models makes it unlikely that this photon could have been observed at the claimed energy. 
If the photon event and its 
energy are confirmed and possibly even more photons above 10 TeV have been observed, the photon-pair production process would have to be suppressed by 
mechanisms predicted in extensions of the Standard Model of particle physics. 
We consider the possibilities of photon mixing with a light pseudo-scalar (e.g., axion-like particles; ALPs) in 
the magnetic field of the host galaxy and the Milky Way
and Lorentz invariance violation (LIV). In the case 
of photon-ALP mixing, the 
boost factor would reach values  $\sim10^6$ for photon couplings not ruled out by the CAST experiment, but limited by other
astrophysical constraints. Viable scenarios would require either 
very efficient mixing in or near to the GRB or that the largest part of the total luminosity is radiated at TeV energies, different from previous GRB afterglows. 
In the 
case of LIV, required boost factors are achievable
for a LIV breaking energy scale $\lesssim 2\times 10^{29}$~eV
($\lesssim 4\times 10^{21}$~eV)
for the linear (quadratic) modification of the dispersion 
relation. A more simple explanation would be
a misidentification of a charged cosmic-ray air shower.}

\begin{document} \maketitle
\section{\label{sec:intro}Introduction}
The exceptionally bright GRB221009A was registered with the BAT instrument
onboard the \textit{Swift} satellite on 2022-10-09 at 14:10:17 UT. The object
was subsequently detected with the XRT and the UVOT (reported in the Gamma-ray Coordinates Network Notice GCN 32632\footnote{See \url{https://gcn.gsfc.nasa.gov/}}). Given the exceptional brightness of the object and its proximity to the Galactic plane
($l=52^\circ 57^m 27.8^s$, $b=04^\circ 19^\prime 17.9^{\prime\prime}$),
the object was initially considered to be a Galactic X-ray transient (Swift J1913.1+1946). The earlier trigger at 13:16:59.00 UT (in the following $t_0$) 
of the Gamma-ray burst monitor (GBM,
GCN 32636) and the subsequent detection with the large-area telescope (LAT, GCN 32637) at 14:17:05.99 UT on board
the \textit{Fermi} satellite made clear that the object is the brightest
gamma-ray burst event detected so far. The emission was sufficiently
bright that photons were detected with the LAT   $\sim25$~ksec after the burst occurred. The
most energetic photon
with $E_\gamma=99.3$~GeV (detected at $t_0+240~\mathrm{s}$) 
is the photon with the highest energy detected from any GRB with \textit{Fermi}~LAT\footnote{the previous record holder was observed with 
$E_\gamma=95$~GeV
from  GRB130427A at $z=0.34$.} (GCN 32658). The energy spectrum above 100~MeV is
characterised with a power law with photon index $-1.87\pm 0.04$ with
an integrated flux between 100 MeV and 1 GeV of 
$(6.2\pm0.4)\times10^{-3}~\mathrm{cm^{-2}~s^{-1}}$. The quoted
flux is 
averaged in the time interval 
from $t_0+200$~s  to $t_0+800$s.   
The temporal structure of the GBM light curve indicates that the first
maximum  was followed by a brighter emission period with multiple maxima 
with a duration $t_{90}\approx 327~\mathrm{s}$ (GCN 32642). The 
fluence of
the second maximum at $0.03~\mathrm{erg/cm^2}$ 
is affected by saturation. The peak photon flux at $t_0+238~\mathrm{s}$ is
$(2385\pm3)~\mathrm{ph~cm^{-2}~s^{-1}}$.

Given the exceptional brightness at X-rays and high energy gamma rays, it
is expected that the object is rather nearby. The first red-shift estimate
obtained at $t_0+11.55~\mathrm{hrs}$ from a spectrum taken 
with the X-shooter at VLT (GCN 32648)
is based upon the detection of absorption lines (CaII, CaI) at $z=0.151$ ($d_L=753~\mathrm{Mpc}$). This
measurement is confirmed with the GTC (GCN 32686: $z=0.1505$)  and provides an estimate of the isotropic energy
release using the GBM fluence of $2\times 10^{54}~\mathrm{erg}\approx 1 M_\odot c^2$. According to the  preliminary classification of the  object (GCN 32686), GRB221009A is   a type II (collapsar) burst \cite{2002A&A...390...81A,2007AdSpR..40.1186Z} or 
a binary-driven hypernova (GCN 32780) \citep{2012ApJ...758L...7R}.

\section{\label{sec:obs} Origin of the 18 TeV photon}
The observational data obtained with various space-based X-ray and gamma-ray instruments will provide important information on the origin of the emission and
the nature of the object. However, in the following, we will focus our analysis
on the  
very high energy emission that was observed with the LHAASO experiment.\footnote{In this context, two other
independent studies related to the optical depth \cite{2022arXiv221005659G} and violation of Lorentz invariance violation \cite{2022arXiv221006338L} have 
appeared as pre-prints
} The LHAASO collaboration reported in GCN 32677 the detection of the GRB both with the water Cherenkov Detector Array (WCDA)
(significance of $100~\sigma$) and the 
larger air shower detector KM2A (significance of $10~\sigma$) within $t_0$ and $t_0+2000~\mathrm{s}$. The total number of 
photons above 500~GeV reported is $\ge 5000$ reaching up to a maximum photon energy 
of $18$~TeV.  In the following, we assume the detection 
of the most energetic photon with KM2A which features a relative energy resolution  of $\sigma_E/E_\gamma \approx 40~\%$ \cite{Ma:2022aau}. The object was observed at 
a zenith angle distance  from $30^\circ$ to $35^\circ$. In Fig.~\ref{fig:sed}, the spectral energy distribution (SED)
of GRB221009A as measured with \textit{Fermi}-LAT is extra-polated
to higher energies, where the observed flux is affected by photon-photon pair production on the extra-galactic background light. 
As a comparison, we have included in Fig.~\ref{fig:sed} 
the SED of the Crab Nebula in 
the same energy range \cite{2022arXiv220311502D}. 
The peak in the SED of GRB221009A is located
beyond the energy range covered with \textit{Fermi}-LAT.
For a peak energy of $\approx 100$~GeV, the extra-polated 
and time-average power  received from
GRB221009A from $t_0+200~\mathrm{s}\ldots 800~\mathrm{s}$ is approximately a factor $10^4$ larger than the power received  from the Crab Nebula. 

\subsection{Absorption in the extra-galactic background light}
The process of photon-photon pair-production ($\gamma+\gamma \rightarrow e^+ + e^-$) has been known for 
a long time \cite{1967PhRv..155.1408G} to suppress the apparent brightness of gamma-rays 
with energies $E_\gamma\gtrsim 100$~GeV. 

The resulting optical depth $\tau$ of a photon
of observed energy $E_\gamma$ relates closely to the 
number of photons per volume and energy interval 
present in the inter-galactic medium at wavelength $\lambda$ 
with $\lambda\approx 1.2~\mu\mathrm{m} (E_\gamma/\mathrm{TeV})$. Since
the EBL is difficult to detect against dominating foreground emission
of the interplanetary and interstellar medium, the optical depth
is calculated under the assumption of an underlying model for the
EBL. Even though the various models and both direct and indirect
estimates of the EBL have been converging in the past decades for
the wavelength range close to the optical and near-infrared, the
uncertainties at larger wavelengths remain substantial.

In Tab.~\ref{tab:events2}, we list the optical depth values 
for a photon of nominal energy $E_\gamma=18$ TeV and $E_\gamma - \sigma_E\approx 10$~TeV  
detected from a red shift $z=0.1505$ using a selection of models for the 
 EBL.\footnote{We use the public repository
of collected tables at \href{github}{https://github.com/me-manu/ebltable}.}
For a given optical depth $\tau$, the observed flux is attenuated by a factor 
$\exp(-\tau)$ which in this case ranges from
$\exp(-27.1)=1.7\times 10^{-12}$ for the upper bound model of Ref.~\cite{2011MNRAS.410.2556D} at $E_\gamma=18$~TeV to $\exp(-4.5)=10^{-3}$ for the lower bound model of Ref.~\cite{2011MNRAS.410.2556D}
 at $E_\gamma=10$~TeV. 
The  photon density of the 
EBL in the mid infra-red at wave-lengths from $\lambda=5~\mu$m to $\lambda=20~\mu$m is most relevant for the estimated optical depth
at gamma-ray energies above 10~TeV. 
The accuracy of the models suffer however from uncertainties on
the amount of stellar light re-processed by polycyclic aromatic hydrocarbon (PAH)
molecules present in the galaxies.
\begin{table*}[h!]
   \label{tab:events2}
    \caption{
    For a set of models  for the extra-galactic background light (EBL;
K\&D2010: \cite{2010A&A...515A..19K}, 
Fi2010: \cite{2010ApJ...712..238F},    
Gi2012(f): \cite{2012MNRAS.422.3189G},
Do2011($\pm)$: \cite{2011MNRAS.410.2556D}, 
Fr2008: \cite{2008A&A...487..837F}, 
SL2021: \cite{2021MNRAS.507.5144S})
we list the resulting optical depth values $\tau$ for
the nominal photon energy of $18~\mathrm{TeV}$ ($\tau_{18}$) and for a lower bound
of $10$~TeV  ($\tau_{10}$) 
within the range of a relative energy resolution of
$\approx 40~\%$.
    Expected number of photons for $\zeta=5$, $t_1=200$~s, 
    above $500$~GeV ($18$~TeV): $N_{\gamma,0.5}$ ($N_{\gamma,18}$) and required coupling to 
    ALPs $g_{a\gamma}$ in units of $10^{-11}$~GeV$^{-1}$,
    as well as (maximum) energy for Lorentz invariance breaking 
    ($M_{1,2}$ for linear and quadratic modifications of the
    dispersion relation) in units of Planck mass $M_\mathrm{Pl}=(\hbar c^5/G)^{1/2}\approx 1.22\times 10^{28}~\mathrm{eV}$. 
    }
    \begin{center}
\begin{tabular}{|l|cccccc|}
\hline
 EBL Model & $\tau_{18}$ & $\tau_{10}$ & $N_{\gamma,0.5}$ & $N_{\gamma,18}$ & $g_{a\gamma}$ & $M_1(M_2)$ \\
 \hline
K\&D2010         &    9.4 &    4.5 &   6700     &    $1$ &     -&    - ( -)  \\
Fi2010           &   10.0 &    6.0 &       4162 &    
$0.9$ &     -&     - (-)  \\
Gi2012         &   13.3 &    5.4 &       4500 &    
$2\times 10^{-2}$ &     0.58 &     10.4 ( $2.6\times 10^{-7}$)  \\
Do2011- &   13.5 &    4.4 &       5800 &    
$1\times 10^{-2}$ &     0.58&     11.3 ( $2.8\times 10^{-7}$)  \\
Gi2012f   &   13.9 &    5.6 &      5603 &    
$1\times 10^{-2}$ &     0.58&     10.1 ( $2.6\times 10^{-7}$)  \\
Fr2008    &   18.3 &    6.8 &       5000 &    
$9\times 10^{-5}$ &     0.59&     8.4 ( $2.4\times 10^{-7}$)  \\
SL2021   &   19.1 &    6.9 &       5200 &    
$4\times 10^{-5}$ &     0.59&     8.4 ( $2.4\times 10^{-7}$)  \\
Do2011       &   19.2 &    6.1 &       4600 &    
$3\times 10^{-5}$ &     0.59&     9.1 ( $2.5\times 10^{-7}$)  \\
Do2011+ &   27.1 &    7.8 &       4000 &   
$7\times 10^{-9}$ &     0.59&     7.5 ( $2.1 \times 10^{-7}$)  \\
\hline
\end{tabular}
\end{center}
\end{table*}

\begin{figure}[ht!]
\begin{center}
\includegraphics[width=1\linewidth]{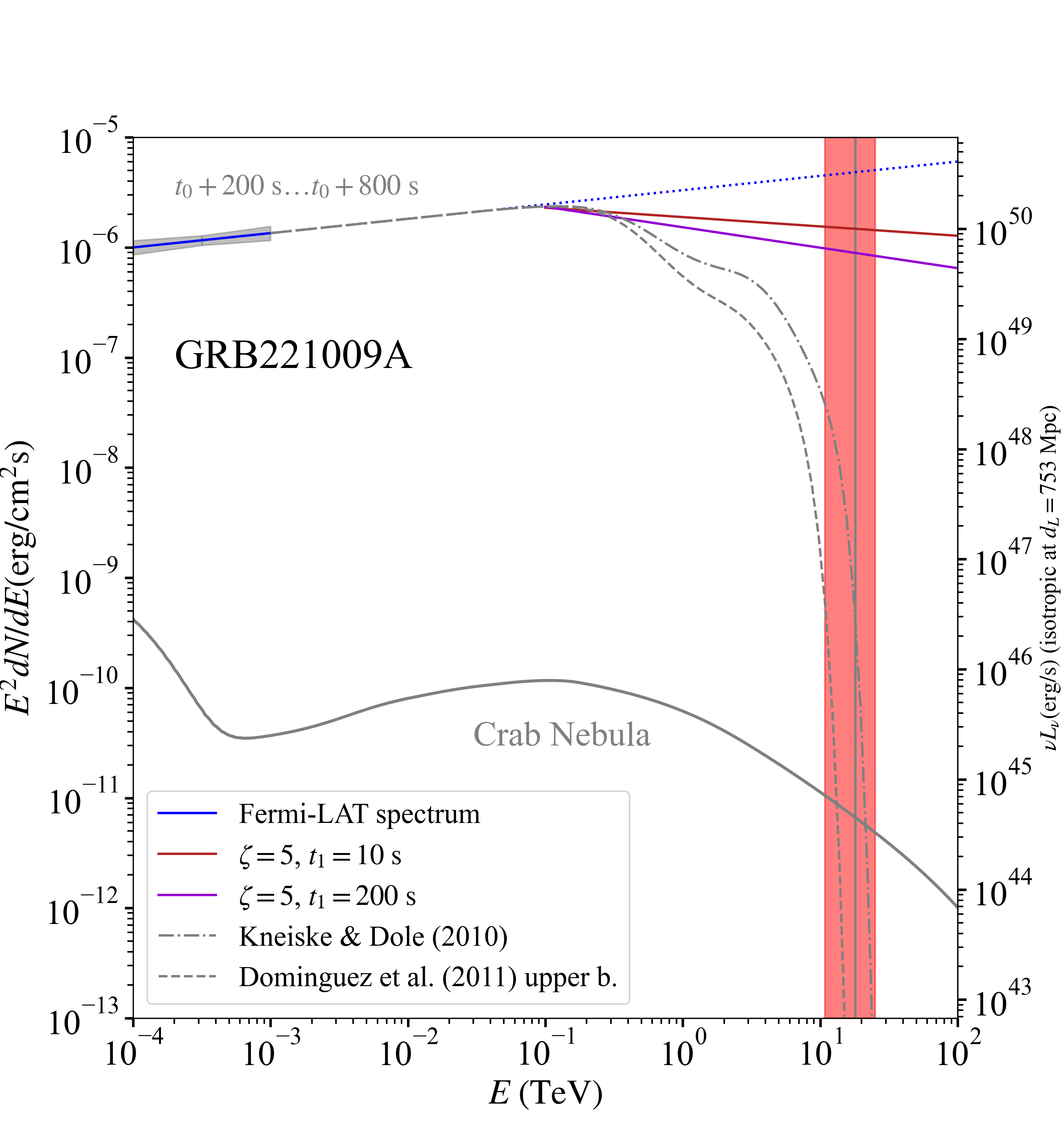}
\end{center}
\caption{Spectral energy distribution (SED) of the high and very high energy
emission of GRB221009A. The bow-tie marks the estimated
confidence region of the \textit{Fermi}-LAT 
time averaged spectrum ($t_0+200~\mathrm{s}\ldots t_0+800~\mathrm{s}$) and its extrapolation to the highest energies (dotted line). The dash and dash-dotted lines indicate the resulting observable flux when assuming two EBL models which provide the largest and smallest optical depth. For the
quantitative analysis, we consider a time-dependent model (see text for details) with two free parameters $\zeta=F_{VHE}/F_{X}$
and onset time of the afterglow $t_1$. Two exemplary curves are shown
in red and magenta.
The vertical grey line  marks the maximum energy $E_\gamma$, while
the shaded region indicates the  
relative energy resolution of $40~\%$.
As a comparison, the SED of the Crab Nebula is overlaid \cite{2022arXiv220311502D}. }
\label{fig:sed}
\end{figure}

The effect of absorption leads to a strong suppression of the observable flux
at TeV energies. In Fig.~\ref{fig:sed}, we show for orientation the 
flux measurement of GRB221009A during the time-interval from $t_0+200~\mathrm{s}$ 
to $t_0+800$~s obtained with \textit{Fermi}~LAT (GCN 32658). The uncertainty on
the flux and the photon index constrain a bow-tie shaped region in the spectral
energy distribution (SED) shown in Fig.~\ref{fig:sed}. The effect
of photon-photon pair-production leads to a noticeable suppression 
of the apparent brightness at energies exceeding $\sim 100~\mathrm{GeV}$. The  observable flux at $E_\gamma=18~\mathrm{TeV}$
is substantially attenuated (in the deep exponential suppression). 
For the models for the EBL with smaller optical depth \citep{2010A&A...515A..19K,2010ApJ...712..238F},
the extrapolated and attenuated flux at 18~TeV 
reaches $\approx 100$ times the
flux of the Crab Nebula. For the model of Ref.~\cite{2011MNRAS.410.2556D}, the flux at 
the same energy would be less than $10^{-4}$
of the Crab Nebula flux. 

\subsection{Expected photon and background counts}
In the following, we estimate the number of photons ($N_\gamma$) expected to be detected 
with LHAASO KM2A within 2000~s after the 
trigger time. Instead of extrapolating the \textit{Fermi}-LAT flux,
we assume that similar to previous GRBs with VHE afterglows, 
the VHE flux in the afterglow phase is proportional to the 
soft X-ray flux. This way, we are only left with two unknown parameters, the
time of the onset of the afterglow emission 
(in the following: $t_1$) and the relative normalisation between the X-ray and VHE band ($\zeta$).

Since the afterglows of GRB221009A and  
 GRB190114C share similarities (e.g. the 
 afterglow emission starts a few minutes after the trigger and
 the observed X-ray light curve follows a power law), we use the 
 time-dependent VHE afterglow spectrum  measured with the
 MAGIC Cherenkov telescopes as a template for
 GRB221009A \citep{2019Natur.575..455M,2019Natur.575..459M}.
 We therefore consider the following assumptions based on observations to estimate the
 flux and spectral shape for the VHE afterglow of GRB221009A:
 \begin{enumerate}
  \item The VHE-afterglow 
  emission starts at a time $t_0+t_1$ after the 
     trigger time $t_0$  with $t_1\approx 10~\mathrm{s}\ldots 200~\mathrm{s}$.
  \item  The ratio $\zeta=0.1\ldots 5$  of the VHE energy flux and the soft X-ray energy flux measured remains constant during the afterglow phase.
  \item The VHE energy spectrum softens during the afterglow. 
 \end{enumerate}
The second assumption follows from the observation that previous X-ray and VHE-afterglow light curves share a similar power-law behaviour (for GRB\-180720B see \citep{2019Natur.575..464A} and for GRB\-190114C see \citep{2019Natur.575..459M}).
Observations of GRB221009A with the \textit{Swift}-XRT started 
about an hour after the trigger and continued up to about a week later.
The X-ray afterglow emission $F_X(t)$ follows a power law in time such that $F_X(t)\propto t^{-\alpha}$ with
$\alpha=1.58$ (GCN32802), similar to the value found for 
GRB190114C. 

  The VHE energy spectrum of GRB190114C was 
  softening during the first 2000~s such that the photon index after correcting for absorption on the EBL was changing from an initial value of $\Gamma(t_0+100~\mathrm{s})\approx-1.9$ to $\Gamma(t_0+2000~\mathrm{s})\approx -3$ \citep{2019Natur.575..459M}.
The differential flux $dF = \phi(E,t)~\mathrm{d}E\mathrm{d}t$ for the afterglow 
starting at $t_1$ is therefore
\begin{equation}
 \phi(E,t) = \phi_0(t) \left(\frac{t}{t_n}\right)^{-\alpha} 
 \left(\frac{E}{E_0}\right)^{-\Gamma(t)},
\end{equation}
where $t_n$ is an arbitrary 
time that determines the 
normalisation.

We choose $\phi_0(t)$ such that the energy flux
integrated in the VHE band from $E_1=1~\mathrm{TeV}$ to
$E_2=10~\mathrm{TeV}$
\begin{equation}
  F(t) = \int\limits_{E_1}^{E_2} \mathrm{d}E\,E\phi(E,t) 
\end{equation} 
follows a power law in time
\begin{equation}
  F(t) = F_0 \left(\frac{t}{t_n}\right)^{-\alpha},
\end{equation}
with $\alpha=1.5$, similar to the XRT light curve and 
$F_0$ chosen such that for all times $t>t_1$
\begin{equation}
 F(t) = \zeta~F_X(t),
\end{equation}
where $F_X$ is the energy flux in the XRT band.
The relative normalisation $\zeta$ has been found to be close to unity in previous observations. The value of $\zeta$ is 
 linked to the Comptonization parameter $Y$  often used in the context of 
 Synchrotron Self-Compton models (see e.g.~\citep{2022MNRAS.512.2142Y}), 
 where naturally values close to unity are expected. We explore the
 range from $\zeta=0.1$ to $\zeta=5$ (see also Fig.~\ref{fig:sed} where
 the VHE flux for $\zeta=5$ is shown).

The photon index $\Gamma(t)$ and its dependence on time
is parameterised with 
\begin{equation}
 \Gamma(t)=\Gamma_0 + a\log(t/t_n),
\end{equation}
where $\Gamma_0=1.5$ and $a=0.44$
for $t_n=200~\mathrm{s}$ are chosen to match the measured
VHE spectrum of the afterglow from GRB190114C \citep{2019Natur.575..455M}. 
The gradual softening of the spectrum from $\Gamma=1.5$ to $\Gamma\approx 3$
after about one hour of afterglow evolution  corresponds 
to a shift of the peak in the spectral energy distribution 
from beyond $10~\mathrm{TeV}$ to less than $1~\mathrm{TeV}$ during 
the observation time. The softening of the VHE spectrum has not been observed
for GRB180720B \citep{2019Natur.575..464A} and GRB190829A \citep{2021Sci...372.1081H} which were however observed at a later phase of the
afterglow.

Any observation of the energy spectrum from a GRB 
will necessarily 
be an average of $\phi(E,t)$ 
over time\footnote{Since we require the time-average spectrum
to be a power law, we actually average $\langle \log \phi\rangle_t$}. Therefore, we consider the 
time-average differential spectrum:
\begin{equation}
\phi(E) = \exp(\langle \log \phi_0(t)\rangle - 
               \alpha \langle \log (t/t_n)\rangle) 
               \left(\frac{E}{E_0}\right)^{-\langle \Gamma\rangle},
\end{equation}
and 
\begin{equation}
 \langle \Gamma \rangle = \Gamma_0 + a \langle \log(t/t_n)\rangle.  
\end{equation}
The time average of $\log (t/t_n)$ between
$t_1$ and $t_1 +\Delta t$ is given by
\begin{equation}
 \langle \log(t/t_n)\rangle = \frac{t_1}{\Delta t} \log\left(1+\frac{\Delta t}{t_1}\right) + \log\left(\frac{t_1+\Delta t}{t_n}\right) - 1.
\end{equation}
With $F_0=\zeta F_X(t=t_n)$ we can re-write 
\begin{eqnarray}
\phi_0(t) = \zeta \frac{F_X(t)}{E_0^2 \int \mathrm{d}x\, x^{1-\Gamma(t)}},
\end{eqnarray}
where the integration is carried out in the interval from $x=E_1/E_0$ to
$x=E_2/E_0$. 

For $t_n=200~\mathrm{s}$, we find from the fit to the 
\textit{Swift}-XRT light
curve $F_X(t_n)\approx 6\times 10^{-6}~\mathrm{ergs\,cm^{-2}\,s^{-1}}$
(GCN32802).

The  time-averaged energy spectrum (assuming no emission for
$t<t_1$) in the LHAASO energy range is shown in Fig.~\ref{fig:sed}
for $\zeta=5$ and for two different values of $t_1$. 
With a given collection area $A(E)$, 
observation time $\Delta t$, and
the time-averaged photon flux 
$\phi(E)$, we calculate the expected number of photons  
$N_\gamma$ in the
energy interval from [$E_{\gamma}$, $80~\mathrm{TeV}$]:
\begin{equation}
\label{eq:extrapol}
N_\gamma(>E_\gamma) =    \Delta t \, \int\limits_{E_\gamma}^{80~\mathrm{TeV}}  \phi(E) \, A(E) 
e^{-\tau(E)}\, \mathrm{d}E.
\end{equation} 
Even though we are primarily interested to obtain an estimate of the 
photon number expected above 18~TeV with the KM2A, we nevertheless
calculate for consistency the number of photons detectable at lower 
energy with the WCDA of LHAASO.

The 
collection area for the WCDA has been released by the  LHAASO collaboration as supplementary information.\footnote{\url{http://english.ihep.cas.cn/lhaaso/pdl/202110/t20211026_286779.html}} The collection area for the KM2A has been
taken from Fig.~2 from \cite{Ma:2022aau}. 
Even though these estimates
may not be directly applicable  to the particular configuration,
event selection cuts, and 
analysis methods underlying the preliminary LHAASO results, it should
be sufficient to get a useful estimate for the KM2A detector\footnote{We note that the WCDA area 
given in Ref.\cite{Ma:2022aau} is roughly 50 times larger
at $E=1$~TeV than the one presented as supplementary information and used here, while the KM2A areas compare
in a consistent way between the two references}. 
The resulting values for $N_{\gamma,0.5}$ and $N_{\gamma,18}$  are listed in Table~\ref{tab:events2} and confirm that the assumptions are
reasonable as the event numbers for the WCDA are roughly reproduced. While  the values of $N_{\gamma,0.5}$ do not vary between the different
EBL models, the differences between the values found for $N_{\gamma,18}$ are considerable between $\approx 10^{-8}\ldots 1$.

The expected number of background events
$N_\mathrm{cr}$ of misidentified charged
cosmic-ray air showers depends upon the choice of event selection cuts. 
The background rate $R_\mathrm{cr}$ at 18\,TeV for the full KM2A array can be read off Fig.~2 from \cite{2021Sci...373..425L} to be $R_\mathrm{cr}\approx 5$ events/hour. 
Consequently, the expected number of misidentified
events for the $\Delta t=2000~\mathrm{s}$
exposure during the activity of GRB221009A
results in $N_\mathrm{cr} \approx 2.8$.

\subsection{Reducing the gamma-ray opacity with Physics beyond the Standard Model}

The high optical depth values 
for some of the considered EBL models reported in Tab.~\ref{tab:events2} suggest that the observation of gamma rays above energies of tens of TeV for the reported GRB redshift is in these cases very unlikely. 
If the LHAASO event turns out to originate from GRB221009A
at an energy of $18~\mathrm{TeV}$, the effective opacity could be
lowered by either  the oscillation between photons and hypothetical light pseudo-scalar bosons or by Lorentz Invariance Violation (LIV).
We discuss these two scenarios in the following. 

\subsubsection{Photon mixing with light pseudo-scalar (axion-like) particle}
\label{sec:alps}

Light pseudo-scalar bosons, often referred to as axions or axion-like particles (ALPs), are predicted in numerous extensions of the Standard Model~\cite[see, e.g.,][for a review]{2018PrPNP.102...89I} and are plausible candidates for cold dark matter~\cite{Preskill_Wise_Wilczek_1983,Abbott_Sikivie_1983,Dine_Fischler_1983,Arias_2012}. 
Photons and ALPs could convert into each other in the presence of external magnetic fields. The oscillation would lead to a reduction of the opacity as these particles do not undergo pair-production with background photons. 

This process has been studied extensively in connection with gamma-ray observations of blazars and considering different magnetic fields along the line of sight \cite[see, e.g.,][for a recent review]{2022Galax..10...39B}. 
As pointed out recently~\cite{2022arXiv221005659G}, photon-ALP conversions could also be responsible for the observation of the 18\,TeV gamma ray reported by LHAASO. 
Here, we investigate if photon-ALP oscillations could yield the required boost in photon flux to explain the LHAASO observations considering a range of ALP masses $m_a$ and photon-ALP couplings $g_{a\gamma}$. 
The boost is defined as the ratio $P_{\gamma\gamma} / \exp(-\tau)$, where $P_{\gamma\gamma}$ is the photon survival probability, i.e., the probability to observe an emitted gamma ray when ALPs are considered. The standard EBL attenuation is given by $\exp(-\tau)$.

For the astrophysical magnetic fields, we consider a minimal scenario: we include the magnetic field of the host galaxy as well as the magnetic field of the Milky Way. 
For the host, we conservatively assume mixing in the regular component of a plausible galactic magnetic field. We take the component transversal to the propagation direction to be 0.5\,$\mu$G coherent over 10\,kpc~\cite[e.g.,][]{2010ASPC..438..197F}. 
The magnetic field of the Milky Way is described by the regular component of the model in Ref.~\cite{2012ApJ...757...14J}. 
We do not consider mixing in the GRB emission region itself as these parameters are at this stage highly uncertain. 
Including the mixing in the emission region with the parameters used in Ref.~\cite{2011JCAP...02..030M} does not change our results. 

\begin{figure}[htb!]
    \centering
    \includegraphics[width=.99\linewidth]{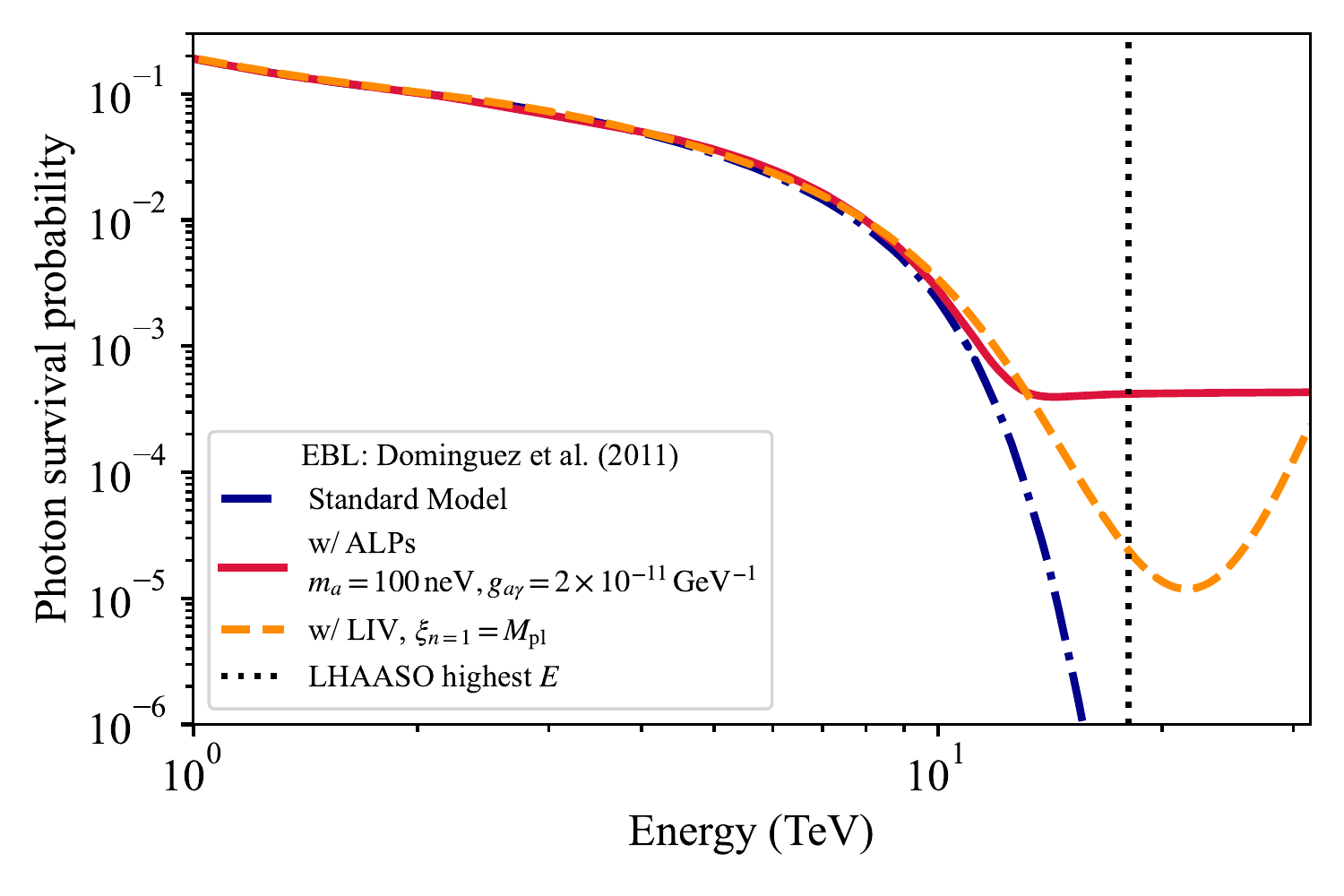}
    \caption{An example for the photon survival probability in the presence of ALPs or LIV as a function of energy. ALPs are produced in the host galaxy of the GRB and reconvert into gamma rays in the Galactic magnetic field. For the chosen magnetic field and ALP parameters, this re-conversion leads to an enhanced transparency above $\sim10\,$TeV. Similarly, the opacity is reduced for the linear LIV effect setting in at energies around the Planck mass $M_\mathrm{pl}$.}
    \label{fig:pgg}
\end{figure}

With these model assumptions, we numerically solve the photon-ALP equations of motion using the transfer-matrix approach implemented in the open-source software package \textsc{gammaALPs}~\cite{2022icrc.confE.557M}.\footnote{\url{https://gammaalps.readthedocs.io/}}
An example is shown in Fig.~\ref{fig:pgg} for $m_a = 10^{-7}\,\mathrm{eV}$, $g_{a\gamma} = 2\times10^{-11}\,\mathrm{GeV}^{-1}$, and the Do2011 model.
Clearly, above $\sim10\,$TeV, the photon flux is greatly enhanced by photon-ALP conversions. 

\begin{figure}[htb!]
    \centering
    \includegraphics[width=.99\linewidth]{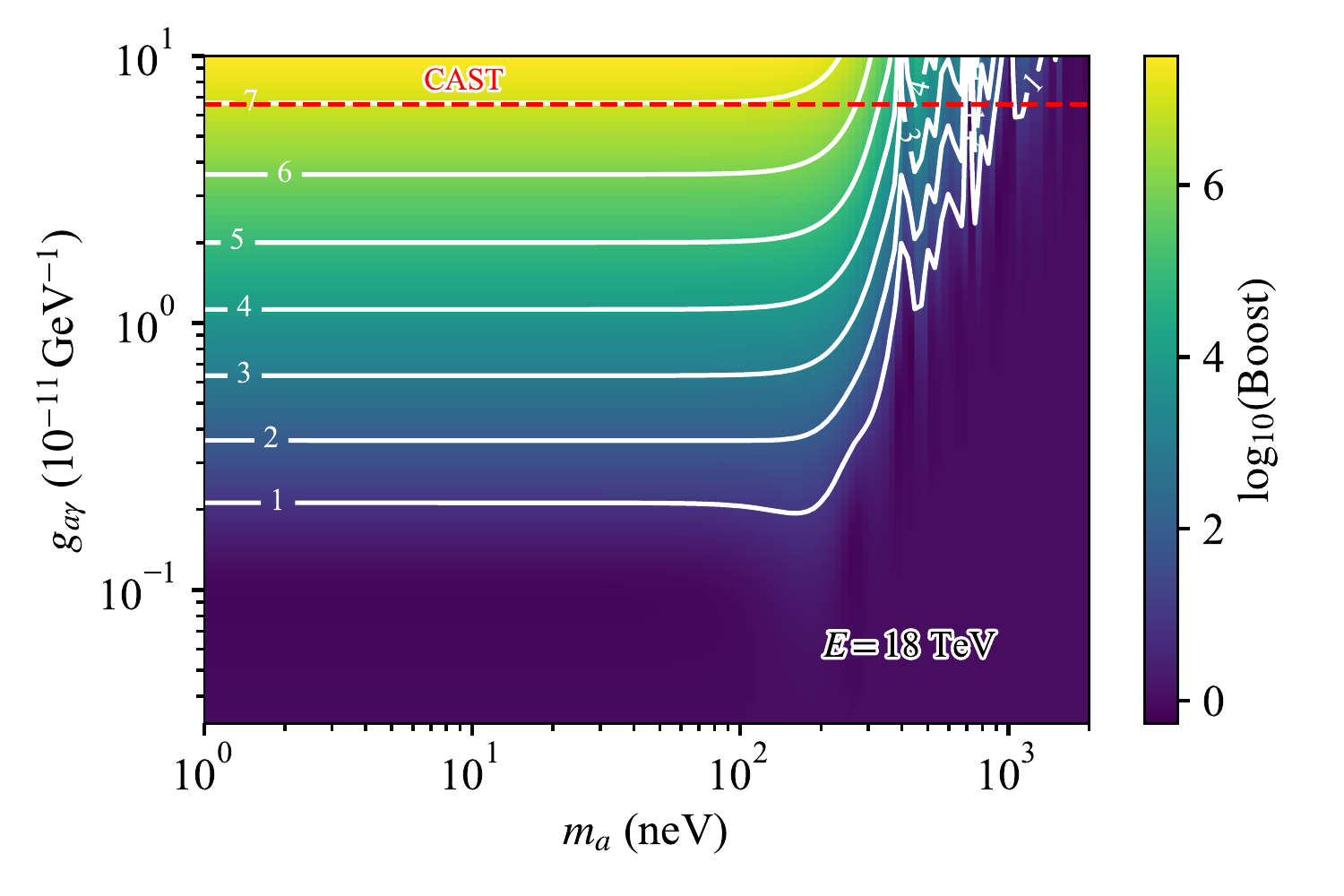}
    \caption{The logarithm of the boost factor of the photon flux over a grid of ALP parameters at $E_\gamma=18\,$TeV. The Do2011 EBL model is assumed. Parameters above the red dashed line are excluded by CAST. }
    \label{fig:boost-grid}
\end{figure}

We now perform the calculation of the boost over a grid of $m_a$ and $g_{a\gamma}$ at an energy of $E_\gamma=18\,$TeV to identify the ALP parameter space (given our model assumptions) which could lead to the required boost to explain the LHAASO observation. 
The result is shown in Fig.~\ref{fig:boost-grid}.
Even for comparatively low coupling values above $2\times10^{-12}\,\mathrm{GeV}^{-1}$ the boost factor is already a factor of 10 for $m_a \lesssim 2\times10^{-7}\,\mathrm{eV}$. 

The values for $g_{a\gamma}$ listed in Tab.~\ref{tab:events2} are estimated by carrying
out the integration in Eqn.~\ref{eq:extrapol} for $\zeta=5$ and 
$t_1=200~\mathrm{s}$ after replacing the optical depth $\tau$
with an effective optical depth $\tau_\mathrm{eff}$ 
such that $\tau_\mathrm{eff}=-\log(P_{\gamma\gamma})$.
 For a fixed value of $m_a=10~\mathrm{neV}$, the equation $N_\gamma(E>18~\mathrm{TeV},g_{a\gamma})=1$
 is then solved for $g_{a\gamma}$. 

For most considered EBL models, the resulting value for $g_{a\gamma}\approx 6~\times 10^{-12}~\mathrm{GeV}^{-1}$ is
not constrained by the latest bound from the CAST experiment~\cite{CAST:2017uph} but is in tension with constraints derived from optical polarisation measurements from magnetic white Dwarfs~\cite{2022PhRvD.105j3034D} ($g_{a\gamma}<5.4\times 10^{-12}~\mathrm{GeV}^{-1}$ at 90~\% c.l.) and from gamma-ray spectra of flat-spectrum radio quasars \citep{2022arXiv220311502D} with
$g_{a\gamma}<5\times 10^{-12}~\mathrm{GeV}^{-1}$ (95~\% c.l.). 

\begin{figure}[ht!]
\centering
\includegraphics[width=.98\linewidth]{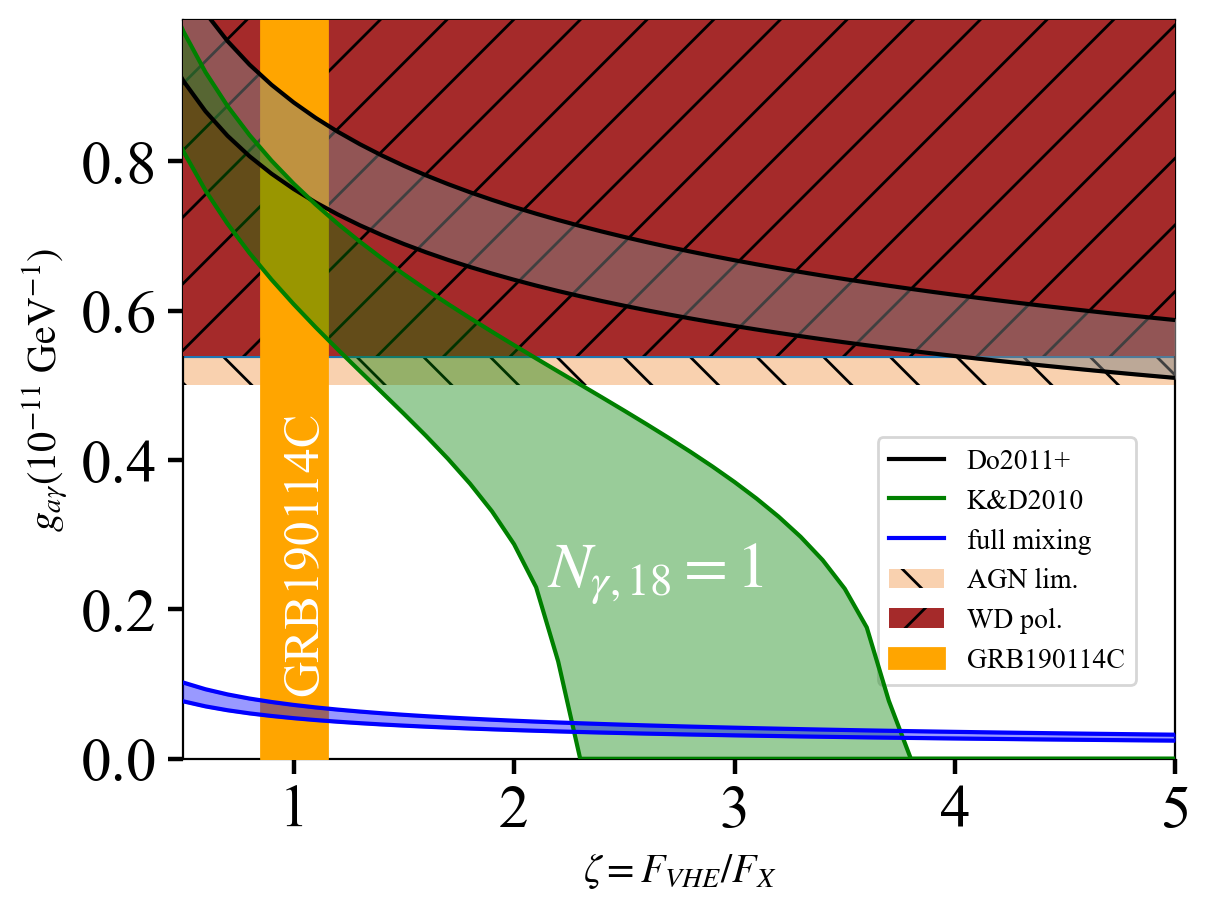}
\caption{
Required photon-ALPs coupling 
 $g_{a\gamma}$ which sufficiently lowers
 the optical depth for a given EBL model
 to achieve the number of photons $N_{\gamma,18}=1$ as a function
 of $\zeta$. Two separate EBL models are considered which mark 
 the extreme values of optical depth (DOM2011+ and K\&D2010 EBL, see also Table~\ref{tab:events2} for more details).
  For any fixed value of $\zeta$, the
resulting band marks the coupling values for which $N_{\gamma,18}$ is unity. The upper and lower bounds relate to the choice of $t_1=200~\mathrm{s}$ and
$t_1=10~\mathrm{s}$ respectively. The upper bounds on allowed values
of $g_{a\gamma}$ are marked as hatched regions for $m_a=10~\mathrm{neV}$.\label{fig:plot_alps} The full 
mixing scenario is independent of the EBL: Here, a maximum mixing in or near the GRB with subsequent complete absorption of photons in the EBL is assumed. The vertical orange region indicates the typical values of $\zeta$ found in previous GRBs with VHE afterglows.}
\end{figure}

In Fig.~\ref{fig:plot_alps}, we compare the favoured range of values of 
coupling $g_{a\gamma}$ for two representative EBL models with the 
existing bounds mentioned above. For the EBL model with a
comparably low optical depth $\tau_{18}\approx 10$, there is no need to invoke any photon-ALPs mixing for values of $\zeta \gtrsim 2.5$.
For models with larger optical depth $\tau_{18}\approx 20$, 
the required value for $g_{a\gamma}$ exceed the bounds mentioned above unless $\zeta\gtrsim 5$. 
If GRB221009A behaves similarly to previous GRBs with VHE-afterglow emission, the typical 
value of $\zeta \approx 1$ would require coupling values for all
considered EBL models that are in 
tension with existing bounds. 

A consistent scenario with $\zeta\approx 1$ and couplings smaller than
the upper bounds for $g_{a\gamma}$ is achievable only for
efficient photon-ALPs mixing in the source.  At this point, we can not exclude such a scenario.
In order to explore the resulting parameters, we assume for simplicity full mixing in the source with subsequent absorption by the EBL such that
a pure ALPs beam is re-converted into photons in the magnetic
field of the Milky Way. The resulting range of couplings required 
is indicated as a blue band in Fig.~\ref{fig:plot_alps}. It is comforting to see that both requirements on $\zeta \approx 1$ and $g_{a\gamma}<5\times10^{-12}~\mathrm{GeV}^{-1}$ can be fulfilled simultaneously. 

\subsubsection{Photon opacity in LIV scenarios}
\label{sec:liv}

Theories of quantum gravity commonly predict LIV~\cite[see, e.g.,][]{2022PrPNP.12503948A} which would have numerous implications for astrophysical observations~\cite[e.g.,][for a review]{2020Symm...12.1232M}.
Here, we largely follow the reviews in Refs.~\cite{2020Symm...12.1232M,2022Galax..10...39B}.
Most importantly in the present context, certain effective field theories predict a modified dispersion relation for photons (and electrons) to leading order $n$.
In the sub-luminal case, the photon velocity (which now depends on the photon energy) is smaller than the speed of light, $c$. In this case, the modified dispersion relation leads to a modification of the pair production threshold. For gamma rays of energy $E_\gamma$ and background photons of energy $\epsilon$ it now reads, 
\begin{equation}
    E_\gamma'\epsilon' \geqslant \frac{2\big(m_ec^2\big)^2}{1-\cos\theta'} + |\xi_n| \frac{E_\gamma^{n+2}}{4M^n},
    \label{eq:ppthr}
\end{equation}
where $m_e$ is the mass of the electron and $\theta$ is the angle between the photon momenta. 
\begin{figure}[htb]
    \centering
    \includegraphics[width=.99\linewidth]{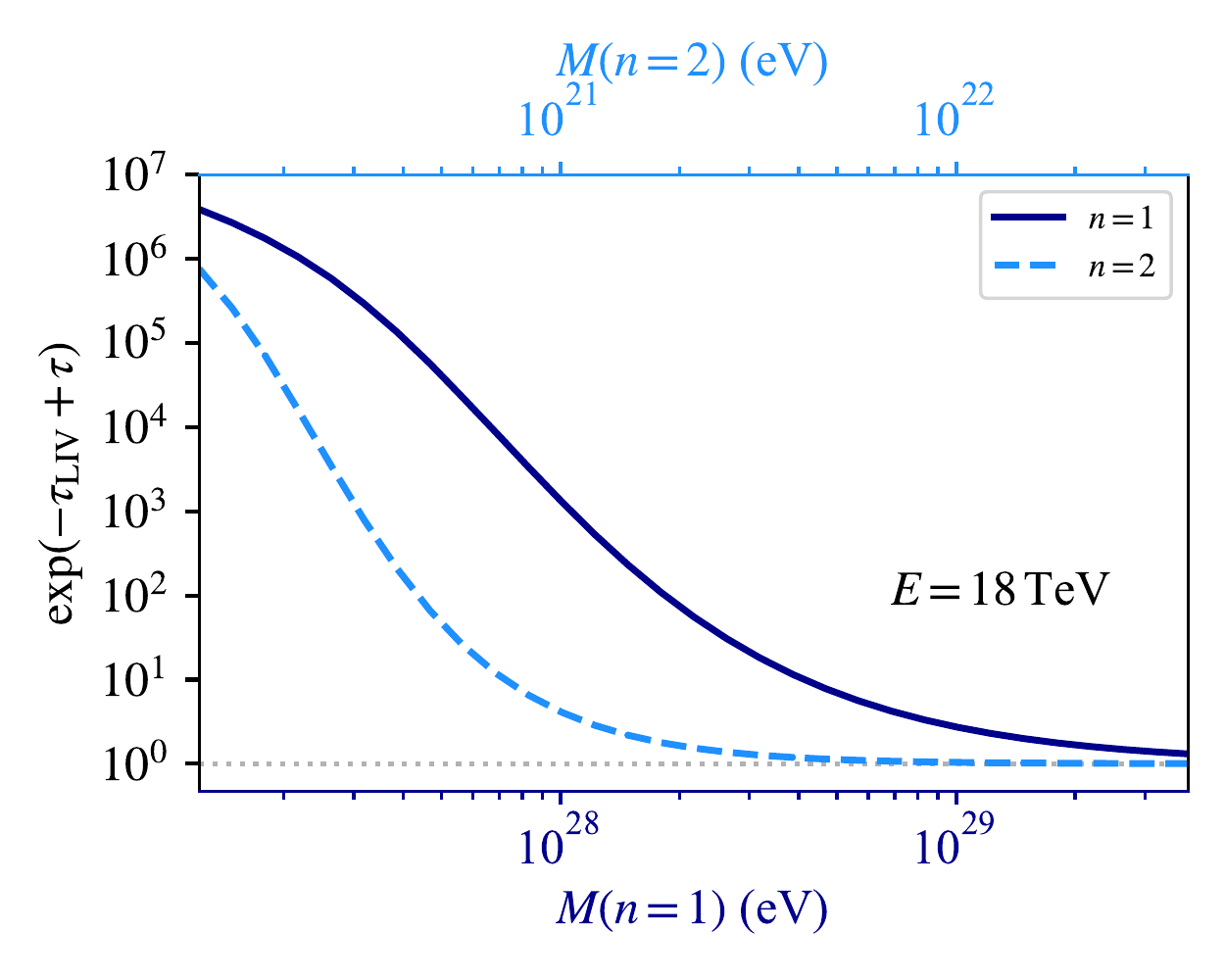}
    \caption{The boost factor for LIV of first (lower $x$ axis) and second order (upper $x$ axis) as a function of the LIV energy scale at 18\,TeV. The Do2011 EBL model is assumed. }
    \label{fig:boost-liv}
\end{figure}

The primed quantities indicate the comoving frame. 
The second term in this equation stems from the modified dispersion relation, where $\xi_n = 1$ if only photons are affected by LIV and $\xi_n = (1 + 2^{-n})^{-1}$ if both electrons and photons are affected. 
The energy scale where LIV becomes important is denoted by $M_n$ and is often given in units of the Planck mass, $M_\mathrm{pl} = \sqrt{\hbar c^5 / G}\sim 1.22\times10^{28}\,$eV. 
As a consequence of the modified threshold, pair production will be suppressed above some gamma-ray energy. 
This is illustrated in Fig.~\ref{fig:pgg} for $M_1 = M_\mathrm{pl}$ for $n=1$ and for LIV only affecting photons  (which we will assume throughout). 
Clearly, the opacity in the presence of LIV, which we denote with $\tau_\mathrm{LIV}$, is reduced above $\sim10\,$TeV and the photon survival probability increases towards one above $20$\,TeV.

\begin{figure}[ht!]
\includegraphics[width=0.49\linewidth]{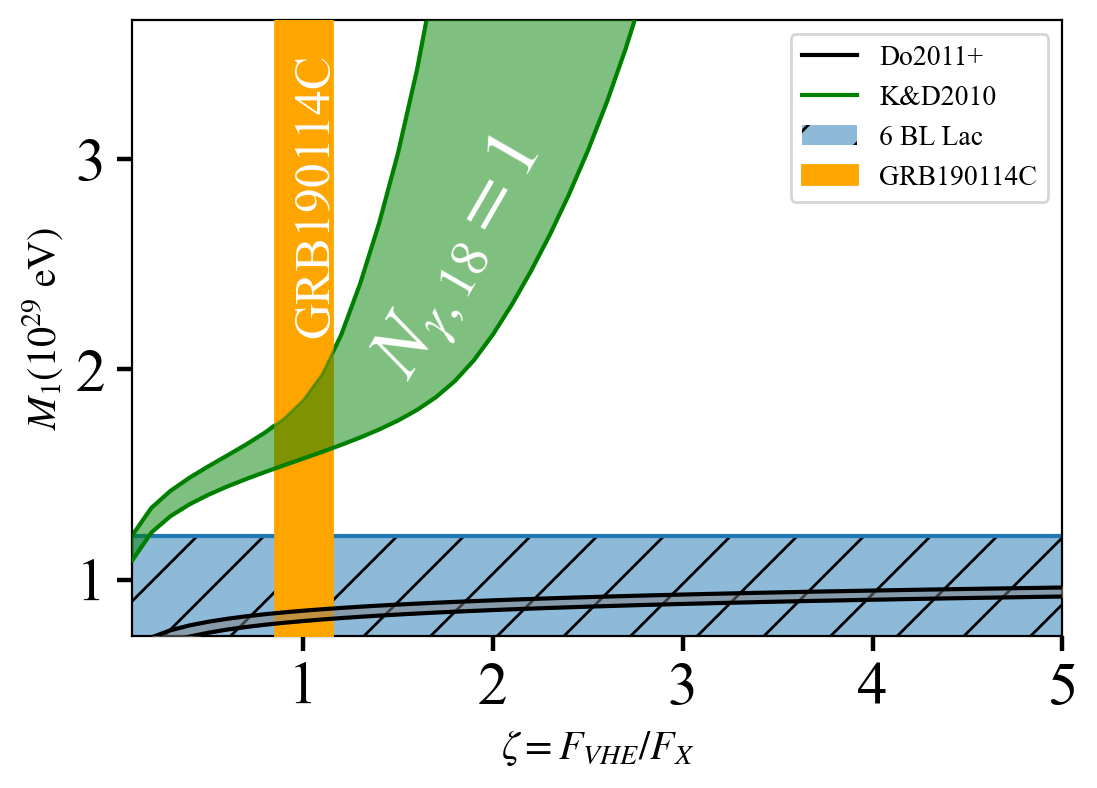}
\includegraphics[width=0.49\linewidth]{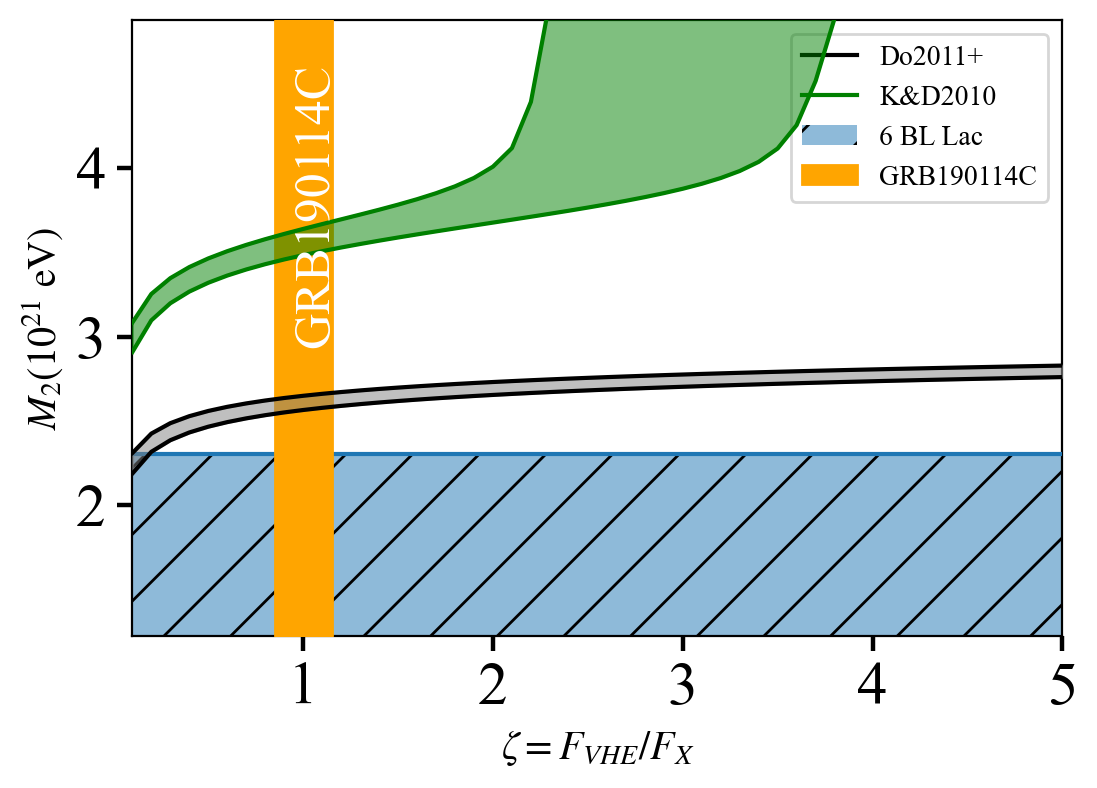}
\caption{Required LIV-breaking energy scale  
which leads to a reduction of the optical depth such that 
$N_{\gamma,18}=1$. The left panel shows the resulting favoured
region for $M_1$ in the linear LIV-breaking scenarios ($n=1$)
while the right panel  contains the favoured regions for $M_2$ in the quadratic ($n=2$) LIV-breaking  scenario. In both cases, the 
favoured range of $M_{1,2}$ depends on the free parameter $\zeta=F_{VHE}/F_X$. The upper and lower bounds of the favoured regions are derived for the onset time $t_1=10~\mathrm{s}$ and $t_1=200~\mathrm{s}$ respectively.
The blue-hatched region marks the exclusion region obtained from
independent observations from gamma-ray spectroscopy. \label{fig:livscan}}
\end{figure}
We now test for which values of the LIV energy scale $M$ we achieve a sufficient boost factor $\exp(-\tau_\mathrm{LIV}+\tau)$ that would explain the observation of an 18\,TeV gamma ray. 
For both the linear $(n=1)$ and quadratic case $(n=2)$, we step through values of $M_1$ 
and $M_2$ respectively. For each value, we calculate $\tau_\mathrm{LIV}$.\footnote{For this, We use the open-source package \textsc{ebltable}, \url{https://github.com/me-manu/ebltable}.}  
The results are shown in Fig.~\ref{fig:boost-liv}. 
Similar to the case of photon-ALPs mixing, we solve the equation $N_\gamma(E_\gamma>18~\mathrm{TeV},M_n)=1$ for $M_1$ and $M_2$ separately resulting
in the values listed in Tab.~\ref{tab:events2}. These values can be considered upper limits
for $M_n$ given that we consider the number of events with 
energies $>18~\mathrm{TeV}$\footnote{A more detailed consideration 
of the number of photons reconstructed at $E_\gamma$ requires detailed
knowledge of the energy resolution that is not available to us.}.

The favoured parameter space for $M_1\lesssim 10^{29}\,$eV 
 is in mild tension
 with
searches for a reduced opacity using observations  of blazars with ground based gamma-ray telescopes. 
In particular, the stacking of observations of multiple sources constrain $M_1 \gtrsim 10^{29}\,$eV~\cite{2019PhRvD..99d3015L}.

The situation is more relaxed for EBL models which predict less absorption as can be seen in Fig.~\ref{fig:livscan} (left panel). The green band
indicates the region with $N_{\gamma,18}=1$ for the EBL model K\&D2010 (similar results are obtained for Fi2010). Here, the required values for $M_1$ even for $\zeta\approx 1$ are above other constraints. In the case of a predominantly 
quadratic LIV-breaking scenario ($n=2$), the optical depth 
is sufficiently reduced for values of $M_2=(2.5\ldots 3.5)\times10^{21}~\mathrm{eV}$ which is not excluded by other constraints (see Fig.~\ref{fig:livscan} right panel).

\section{Conclusions}
The ground-based detection of VHE gamma rays
from GRB221009A provides an exceptional opportunity to 
study the acceleration and emission processes. 
The approximately 5000 photons that are collected above
500~GeV with LHAASO-WCDA from the start of the afterglow 
up to 2000~s after the trigger are exceptional in comparison to previous VHE afterglow observations:
MAGIC detected 877 photons from $t_0+62~\mathrm{s}$ to $t_0+1227~\mathrm{s}$ from GRB190114C \citep{2019Natur.575..455M},  H.E.S.S. detected 119 
photons from GRB180720B during a two-hour observation 
in the late phase of the afterglow (10.1~hours after the trigger time). The afterglow emission from GRB190829A was followed even longer: H.E.S.S. carried out follow-up observations between 4 and 56~hours after the trigger time \citep{2021Sci...372.1081H}. In all three cases, the
VHE afterglows showed a similar temporal behaviour where
the VHE flux $F_{VHE}(t)=\zeta F_{X}(t)$ with $\zeta \lesssim 1$. 

In the approach suggested here, we consider the temporal behaviour of the VHE-afterglow emission of GRB221009A to follow the same pattern and choose the \textit{Swift}-XRT 
light curve of the X-ray afterglow as a proxy to
the VHE-afterglow with $\zeta$ left to vary between $0.1$ and $5$. In the resulting calculation of event numbers, we can reproduce the number of photons detected above 500~GeV with the WCDA detector of LHAASO under the assumption that $\zeta\approx 5$.

The hints for photons detected beyond 10 TeV
for GRB221009A ($z=0.1505$) extend the energy
range  of photons observed during  previous VHE afterglow observations. 
The origin of the claimed highest energy photon detected  
with LHAASO (most likely with the KM2A) at $E_\gamma=18$~TeV is 
at first glance difficult to reconcile with
the large optical depth at that energy due to pair-production with the extra-galactic background light (EBL). 

Among the collection of models for the EBL investigated here (see also Tab.~\ref{tab:events2}), the optical depth is
expected to be between $\tau(18~\mathrm{TeV})=9.4\ldots 27.1$. The 
probability to detect at least one photon for
the extrapolated spectrum calculated with $\zeta=5$ could be therefore as small as 
$7\times 10^{-9}$ but reaches values close to unity for other models.

Possible approaches to suppress pair production include photons mixing
with light ($m\lesssim 10^{-7}$~eV) pseudo-scalar particles (such as ALPs) thus avoiding pair production for a small fraction of the photons that re-convert closer to the observer. 
The resulting values for the mixing $g_{a\gamma}$ depend 
on the presence of magnetic fields close to the source. 
Assuming the EBL model of Ref.~\cite{2011MNRAS.410.2556D} and photon-ALP conversions in the hosting galaxies, we find 
 values $g_{a\gamma}\gtrsim6\times 10^{-12}\,\mathrm{GeV}^{-1}$ for $\zeta=5$ to achieve a sufficiently large boost factor
 in most of the considered scenarios. The resulting values do not depend very much on the choice of the EBL
 model as long as the $\tau_{18}\gtrsim 10$. 
The required value for $g_{a\gamma}$ is in tension with bounds from other astrophysical observations which constrain $g_{a\gamma}\lesssim 5\times 10^{-12}~\mathrm{GeV}^{-1}$. 
For values of $\zeta\approx 1$, the required photon-ALPs mixing is inconsistent with other bounds. However, a modified scenario with maximum mixing of photon and ALPs 
near or inside the source would 
reduce the required  $g_{a\gamma}\lesssim 10^{-12}~\mathrm{GeV}^{-1}$.

Alternatively, breaking of Lorentz invariance  could decrease the gamma-ray opacity. For the quadratic LIV-breaking scenario and $\zeta\gtrsim 0.1$, the required 
values of $M_2$ are larger than bounds from other observations, independent of the choice of the EBL. 
In the linear LIV-breaking scenario, only the scenarios
with a comparably small optical depth from a lower level of the EBL would not require
values of $M_1$ that are inconsistent with other bounds. It will be interesting
to study the arrival time of the TeV photons from GRB221009A to constrain LIV breaking effects independently of the optical depth.

Even if it is exciting to  invoke new physics to explain the
origin of the photon at 18~TeV, it should be mentioned that the
background expectation to detect a mis-identified 
charged cosmic-ray induced air shower
at 18\,TeV is consistent with the actual observation. 

In conclusion, we have demonstrated that the transparency
of the universe could be substantially larger than anticipated 
for photon-photon pair production if photon-ALP mixing is effective or Lorentz invariance violation occurs at sufficiently low energy scales.
Consequently, the observation of bright transients at
large optical depth could in principle
lead to the discovery of anomalous transparency and provide important clues on the nature of the mechanism for the suppression of pair production. 

\section*{Acknowledgments}
We acknowledge the support from the Deutsche Forschungsgemeinschaft (DFG, German Research Foundation) under Germany’s Excellence Strategy – EXC 2121 ``Quantum Universe'' – 390833306.
M.~M.\  acknowledges  support from the European Research Council (ERC) under the European Union’s Horizon 2020 research and innovation program Grant agreement No. 948689 (AxionDM).

\bibliographystyle{JHEP}
\bibliography{grb}
\end{document}